\documentclass[conference,compsoc]{IEEEtran}
\IEEEoverridecommandlockouts
\usepackage{cite}
\usepackage{amsmath,amssymb,amsfonts}
\usepackage{algorithmic}
\usepackage{graphicx}
\usepackage{textcomp}
\usepackage{xcolor}
\usepackage{algorithm}
\usepackage{booktabs}
\usepackage{multirow}
\usepackage{url}
\usepackage{hyperref}
\usepackage{subcaption}
\usepackage{enumitem}
\usepackage{placeins}
\graphicspath{{images/}{figures/}}

\def\BibTeX{{\rm B\kern-.05em{\sc i\kern-.025em b}\kern-.08em
    T\kern-.1667em\lower.7ex\hbox{E}\kern-.125emX}}

\begin{document}

\title{MeshDNS: A Cooperative DNS Resolution Framework for Resource-Constrained IoT Networks\\
}

\author{
\IEEEauthorblockN{Asif Mahbub}
\IEEEauthorblockA{
North South University\\
Dhaka, Bangladesh\\
asif.mahbub01@northsouth.edu
}
\and
\IEEEauthorblockN{Md. Abir Hossain}
\IEEEauthorblockA{
North South University\\
Dhaka, Bangladesh\\
abir.hossain06@northsouth.edu
}
\and
\IEEEauthorblockN{Nabil Bin Hannan}
\IEEEauthorblockA{
North South University\\
Dhaka, Bangladesh\\
nabil.hannan@northsouth.edu
}
}


\maketitle

\begin{abstract}
Domain Name System (DNS) resolution in Internet of Things (IoT) networks presents unique challenges due to resource constraints, unreliable connectivity, and security vulnerabilities. Traditional centralized DNS architectures introduce single points of failure. This paper presents MeshDNS, a cooperative DNS resolution framework designed for resource-constrained IoT environments operating under shared-key admission. MeshDNS employs a decentralized architecture where nodes maintain cache awareness using hash-based summaries and secure cold-cache misses via Ed25519-signed, identical-answer quorum voting. Our implementation on commodity ESP8266 microcontrollers (sub-50~KB usable RAM, 80~MHz) achieves a 0.47~ms warm-cache resolution, outperforming native mDNS baselines (1.39~ms). To secure initial cold-cache misses, MeshDNS trades a predictable $\sim$1.3--1.7\,s cryptographic penalty to successfully isolate Byzantine faults among admitted peers. Assuming a threat model where physical hardware extraction remains out of scope, MeshDNS demonstrates Byzantine fault isolation. We validated the framework via a 5-node physical testbed and discrete-event simulations scaling to 1,000 nodes; the results demonstrate that MeshDNS maintains resilient local name caches for persistent edge telemetry under network churn. Code is available at \href{https://github.com/mahbubasif/MeshDNS-Artifact}{https://github.com/mahbubasif/MeshDNS-Artifact}.
\end{abstract}


\section{Introduction}

The proliferation of Internet of Things (IoT) devices has transformed the landscape of network computing with more than 75 billion connected devices~\cite{iot_forecast_2020}. These resource-constrained devices face challenges in fundamental network operations, including Domain Name System (DNS) resolution—a critical component for internet connectivity that translates human-readable domain names to IP addresses~\cite{mockapetris1987dns}. Furthermore, optimizing the deployment and network hardening of such IoT systems is critical to mitigate the exploitation of vulnerabilities throughout the attack surface~\cite{agmon2019deployment}. The need for lightweight, dynamic security mechanisms is paramount given the severe resource and computational constraints inherent to edge devices~\cite{zhao2022dywcp,sciancalepore2022privacy}.

Traditional DNS architectures rely on hierarchical client-server models with centralized root servers, authoritative nameservers, and recursive resolvers~\cite{dns_root_servers}. While this design has proven effective for conventional internet infrastructure, it introduces several critical limitations for IoT deployments:

\begin{enumerate}
    \item \textbf{Latency bottlenecks}: IoT devices must traverse multiple network hops to reach external DNS servers, introducing resolution delays of 50-200ms per query~\cite{dns_caching_analysis}.
    \item \textbf{Single points of failure}: Centralized DNS infrastructure creates vulnerability to outages, DDoS attacks, and network partitions~\cite{dns_ddos_2016}.
    \item \textbf{Privacy concerns}: External DNS queries expose device behavior and network topology to third-party resolvers~\cite{iot_dns_privacy}.
    \item \textbf{Scalability limitations}: DNS server load increases linearly with device count, creating performance degradation in large-scale deployments~\cite{dns_scalability}.
    \item \textbf{Energy overhead}: Radio transmission for DNS queries consumes significant battery power, with DNS traffic accounting for 15-20\% of total energy consumption in wireless sensor networks~\cite{energy_dns_wsn}.
\end{enumerate}

Compounding these architectural limitations, modern edge frameworks (e.g., Matter~\cite{csa_matter_solution}, Home Assistant~\cite{homeassistant_concepts}) are aggressively shifting toward local, autonomous device-to-device communication. Because devices in these environments require stable, long-lived connections to local hubs for persistent telemetry, maintaining resilient local name caches is a mandatory prerequisite for survivability. To address this, recent work has explored decentralized alternatives, including distributed hash tables, blockchain-based systems~\cite{namecoin, ethereum_ens}, and peer-to-peer resolution. However, these approaches typically require computational resources well beyond the capabilities of resource-constrained IoT devices (e.g., 32-64~KB RAM, 80-160~MHz processors). Furthermore, they often lack the lightweight, energy-efficient, and authenticated quorum mechanisms necessary for local IoT environments where compromised nodes may exhibit arbitrary malicious behavior~\cite{byzantine_iot}.

\subsection{Contributions}

To address these gaps, this paper presents MeshDNS, a cooperative DNS framework for single-hop IoT LANs. By framing our inquiry around the practical implementation of local cooperative resolution on edge devices, our work makes the following contributions:

\begin{itemize}
    \item We validate the feasibility of a local cooperative DNS framework on heavily resource-constrained edge devices via a 5-node ESP8266 testbed (50~KB RAM) \cite{espressif_esp8266ex_2023}, explicitly scoping our shared-key threat model to exclude physical hardware extraction (e.g., JTAG/EEPROM dumping).
    \item We design a two-plane cooperative architecture---signed quorum voting for resolver agreement, plus trust-gated gossip for authenticated cache awareness, along with discrete-event scalability experiments to 1000 nodes as a protocol stress bound.
    \item We implement signed quorum voting for per-query DNS resolution and evaluate it with a structured hardware adversarial matrix (varying $f$, Level-1/4 attacks, and Sybil $k{=}3$) plus a calibrated \texttt{SimPy} sweep.
    \item We empirically demonstrate a 0.47~ms warm-cache latency (outperforming a 1.39~ms mDNS baseline). While mDNS lacks cryptographic fault tolerance, MeshDNS uniquely trades a $\sim$1.3--1.7\,s signed cold-quorum penalty for authenticated fault isolation, optimizing for TTL-amortized workloads rather than burst discovery.
\end{itemize}
Overall, our work underscores the practicality and necessity of decentralized, cooperative infrastructure for robust IoT deployments in adversarial and resource-constrained environments.

\section{Related Work}
\label{sec:related}
We review prior research on distributed resolution architectures, coordination mechanisms, and resilience techniques relevant to IoT deployments.
\subsection{DNS and Decentralized Naming Approaches}

The Domain Name System~\cite{mockapetris1987dns} employs a hierarchical distributed database with root servers, authoritative nameservers, and recursive resolvers~\cite{dns_root_servers}. While this design has proven remarkably scalable for the global internet, supporting billions of queries daily~\cite{dns_traffic_analysis}, it introduces inherent centralization that conflicts with the distributed nature of IoT deployments. DNS performance optimization and the logistical challenges of DNSSEC deployment monitoring~\cite{osterweil2009deploying} have been extensively studied, including caching strategies~\cite{dns_caching_analysis}, prefetching mechanisms~\cite{dns_prefetch}, and anycast routing techniques~\cite{dns_anycast}. However, such assumptions do not hold in resource-constrained edge environments, where intermittent connectivity, limited processing capacity, and energy constraints result in increased resolution latency, frequent upstream query failures, and reduced reliability of recursive DNS traversal.


On the other hand, Blockchain-based naming systems have emerged as alternatives to centralized DNS. Namecoin~\cite{namecoin} pioneered blockchain-based domain registration using proof-of-work consensus, while Ethereum Name Service (ENS)~\cite{ethereum_ens} provides smart contract-based name resolution. Blockstack~\cite{blockstack} combines blockchain anchoring with off-chain storage for scalable naming. However, these systems require full blockchain synchronization (tens of gigabytes) and proof-of-work validation—prohibitive for IoT devices with kilobytes of RAM~\cite{blockchain_iot_challenges}. Distributed Hash Table (DHT) based systems, including Chord~\cite{chord}, Kademlia~\cite{kademlia}, and Pastry~\cite{pastry}, enable decentralized key-value storage suitable for name resolution.  However, DHT maintenance protocols generate significant control traffic, and the $\mathcal{O}(\log n)$ lookup complexity introduces latency unsuitable for real-time, resource-constrained IoT applications.

\subsection{Cooperative Caching in Networks}

Cooperative caching has been extensively studied for content delivery networks~\cite{cdn_caching}, and peer-to-peer file sharing. Summary Cache~\cite{summary_cache} introduced bloom filter-based cache coordination, while Cache Digests~\cite{web_cache_cooperation} employed directory-based approaches.

In wireless sensor networks, several cooperative caching schemes have been proposed to reduce energy consumption~\cite{wsn_coop_cache_1, wsn_coop_cache_2}. However, these systems typically assume trusted environments, which is unrealistic in adversarial IoT deployments where compromised devices may inject malicious responses. Therefore, ensuring the correctness in cooperative caching requires mechanisms that tolerate arbitrary node failures. Our work extends cooperative caching with Byzantine fault-tolerant voting mechanisms suitable for adversarial IoT deployments.

\subsection{Resilient and Gossip-Based Coordination Mechanisms}

Byzantine fault tolerance, introduced by Lamport et al.~\cite{byzantine_generals}, addresses consensus in distributed systems with arbitrary (malicious) node failures. Practical Byzantine Fault Tolerance (PBFT)~\cite{pbft} reduced complexity from exponential to polynomial time, enabling practical deployments. Recent work has applied Byzantine fault tolerance to IoT and sensor networks. For example, Shi et al.~\cite{shi2023ms} introduced MS-PTP to protect network timing protocols from Byzantine attacks, while recent edge architectures have successfully extended Byzantine robustness to asynchronous distributed learning environments~\cite{fang2022aflguard}. However, existing BFT protocols typically require $3f+1$ replicas to tolerate $f$ failures, introducing communication overhead unsuitable for battery-powered devices. Voting-based quorum mechanisms have been applied to IoT data aggregation and intrusion detection~\cite{iot_voting_ids}. Our work adapts signed per-query quorum voting specifically for DNS resolution, optimizing for the unique characteristics of name lookup workloads.

Epidemic (gossip) protocols enable efficient information dissemination in large-scale distributed systems~\cite{gossip_survey}. SWIM~\cite{swim} introduced gossip-based failure detection with constant message load, while HyParView~\cite{hyparview} provides scalable peer sampling for unstructured overlays. These decentralized protocols have been successfully adapted for resource-constrained wireless sensor networks, including epidemic-based code propagation and software updates~\cite{gossip_iot_updates} and time-synchronized mesh networking~\cite{gossip_time_sync}. The Trickle algorithm further introduced adaptive propagation rates that significantly reduce transmission overhead~\cite{gossip_iot_updates}.

Our work integrates lightweight voting-based fault tolerance---the resolution plane---with gossip-style summary dissemination for awareness and churn modeling, keeping authoritative cache agreement on the signed voting path.

\subsection{DNS for IoT}

Recent work has addressed DNS challenges specific to IoT deployments. DNS-SD enables service discovery over DNS, widely adopted in IoT protocols like mDNS~\cite{mdns}. However, mDNS operates only on local networks without internet connectivity.

Shafiq et al.~\cite{iot_dns_security} analyzed the distinct characteristics of machine-to-machine cellular traffic, highlighting the unique network burdens and operational patterns imposed by autonomous edge devices. Furthermore, Siby et al.~\cite{iot_dns_privacy} studied the privacy implications of DNS traffic, finding that even with encryption, traffic analysis of DNS queries can expose sensitive device behavior and usage patterns.

In our work, we provide a complete Byzantine-resilient cooperative DNS framework optimized for resource-constrained devices, validated through prototype implementation on commodity IoT hardware.

\section{System Design}
\label{sec:design}

\subsection{Architecture Overview}

\begin{figure}[htbp]
 \centering
\includegraphics[width=0.80\linewidth]{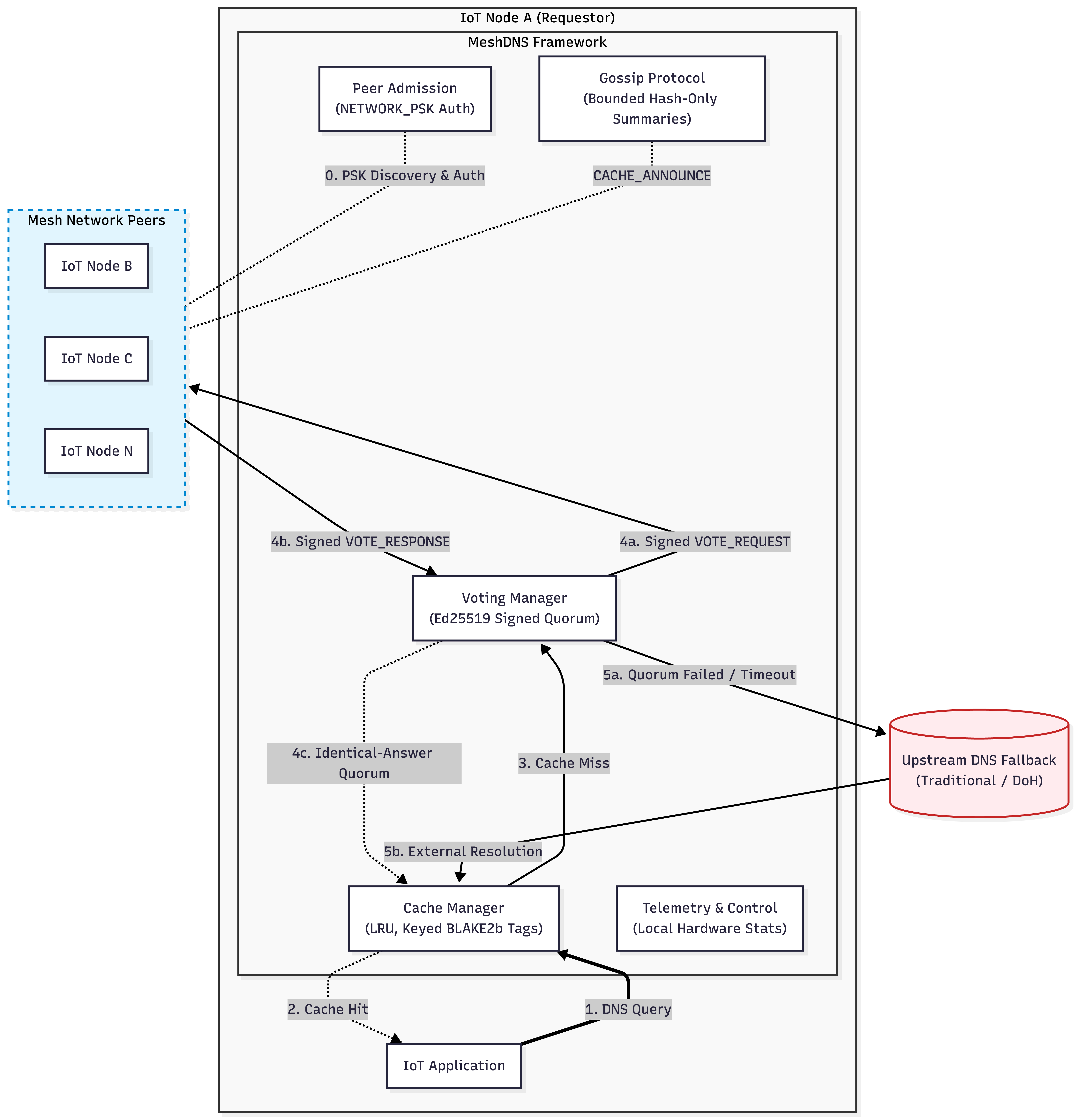}
 \caption{MeshDNS architecture overview illustrating local cache lookup, authenticated peer admission, signed quorum voting, cache-summary gossip, telemetry/control, and upstream DNS fallback.}
\label{fig:architecture}
\end{figure}

MeshDNS employs a local peer-to-peer architecture within a single 802.11 broadcast domain, where nodes maintain local DNS caches and cooperatively resolve queries. Figure~\ref{fig:architecture} illustrates its five core components:

\begin{enumerate}
    \item \textbf{Cache Manager}: Maintains a local LRU cache of DNS records with configurable capacity (default 20 entries on our ESP8266 prototype).
    \item \textbf{Peer Admission}: Authenticates peer discovery using a pre-shared network key and signed discovery messages before admitting a node into the local mesh.
    \item \textbf{Voting Manager}: Implements Byzantine-resilient signed quorum voting for individual DNS resolution rounds.
    \item \textbf{Gossip Protocol}: Periodically announces cache summaries and performs trust-gated, signed \texttt{CACHE\_REQUEST}/\texttt{CACHE\_RESPONSE} prefetch for cache awareness; gossip-only rows are non-authoritative and cannot produce votes.
    \item \textbf{Telemetry and Control}: Exposes hardware-test commands for clearing caches, seeding records, resolving names, collecting statistics, and running cold-quorum experiments.
\end{enumerate}

When a node requires DNS resolution, it first consults its local cache. On a cache miss, it queries authenticated peers via on-demand \texttt{VOTE\_REQUEST}/\texttt{VOTE\_RESPONSE} exchanges. If peers cannot supply a validated quorum answer, the node falls back to traditional hierarchical DNS and caches the result. This three-tier strategy balances low latency (local cache), Byzantine resilience (signed voting), and compatibility (DNS fallback).

\subsection{Network Model and Topology}
Because traditional decentralized networks (e.g., DHTs) rely on multi-hop overlay routing, they suffer from high latency and partition vulnerabilities in poorly connected topologies. MeshDNS explicitly avoids this by restricting its physical operational scope to single-subnet Local Area Networks (LANs), operating instead as a fully connected application-layer overlay mesh.
\begin{figure}[htbp]
 \centering
 \includegraphics[width=0.70\linewidth]{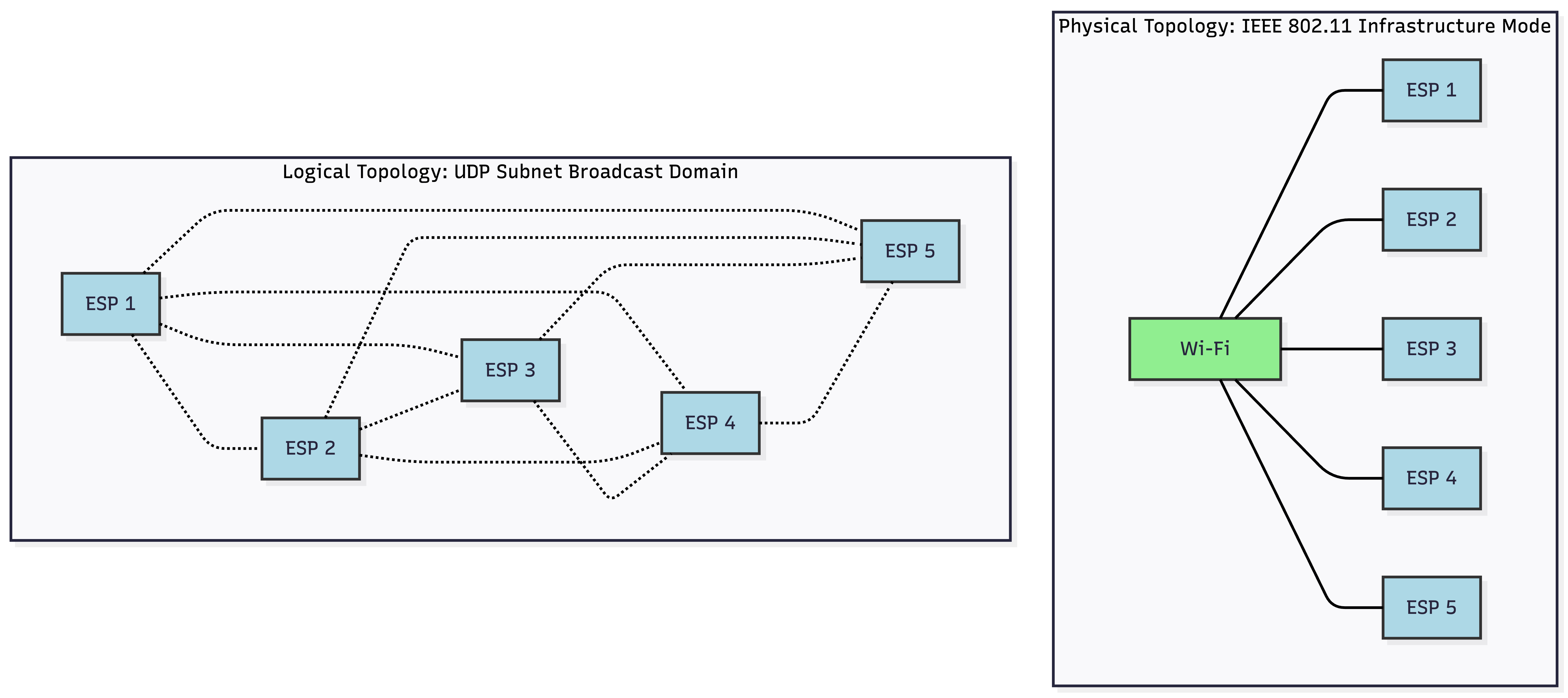}
 \caption{MeshDNS Network Topology. The physical layer (right) utilizes a standard Wi-Fi star topology centered on an Access Point, while the logical protocol layer (left) functions as a fully connected, single-hop UDP broadcast domain.}
\label{fig:topology}
\end{figure}

At the physical layer, nodes operate in standard IEEE 802.11 infrastructure mode, forming a star topology around a central Wi-Fi Access Point (AP) or local router. However, at the logical protocol layer, MeshDNS operates as a fully connected, single-hop broadcast domain. All peer discovery, cache announcements, and Byzantine voting requests are transmitted via UDP subnet broadcasts (e.g., \texttt{192.168.0.255}).

Consequently, there is no multi-hop overlay routing state to maintain. As long as a node maintains a physical connection to the AP, it maintains a direct logical connection to all peers simultaneously, ensuring that quorum voting latency remains tightly bounded by the AP's switching capacity rather than peer-to-peer routing depths.

\subsection{Distributed Cache Management}

Each node maintains a local DNS cache implemented as a hash table with LRU eviction. The cache stores mappings from domain names to IP addresses along with essential metadata. Specifically, each entry contains a plaintext \textit{Domain name} (for resolver fast-path lookup), a 32-byte keyed \textit{Domain tag} (BLAKE2b \cite{rfc7693} with the network pre-shared key, used on the wire and for gossip/voting by hash), an \textit{IP address}, expiry metadata, and a \textit{Reputation} byte (gating local cache hits), distinct from per-peer EWMA trust updated during voting. The application resolver path matches entries by domain string, rejects low-reputation or expired rows, and returns the mapped IP (Appendix~\ref{app:algorithms}); gossip prefetch rows use hash-only placeholders for awareness and deduplication, but the firmware marks them as \texttt{gossip\_prefetch} and never uses them to produce signed peer votes.

\subsection{Byzantine-Resilient Signed Quorum Voting}
\label{sec:voting}

When a cache miss occurs, the node broadcasts a \texttt{VOTE\_REQUEST} containing a 32-byte keyed \textit{Domain tag} ($h$), its Ed25519 \textit{Public key}, a \textit{Timestamp}, and a 64-byte Ed25519 \textit{Signature}~\cite{rfc8032} over $(h \,\|\, \text{timestamp})$. Peers caching the domain reply with a \texttt{VOTE\_RESPONSE} containing $h$, the cached \textit{IP address}, their \textit{Public key}, a \textit{Timestamp}, and a \textit{Signature} over $(h \,\|\, \text{ip} \,\|\, \text{timestamp})$. 

As detailed in Algorithm~\ref{alg:voting}, voting employs a two-phase round. During the 200~ms \texttt{VOTE\_COMMIT\_DELAY\_MS} collect phase, responses are buffered and signatures verified. The firmware strictly enforces this window to capture and penalize late equivocating votes, although Algorithm~\ref{alg:voting} logically illustrates an early-exit upon quorum success for brevity. Counting continues up to \texttt{VOTE\_TIMEOUT\_MS} (3.5~s). The resolver accepts only an identical-answer quorum, penalizing duplicate keys and equivocations. This constitutes a per-query authenticated agreement mechanism rather than PBFT-style global consensus, relying on bounded malicious fractions and timely honest responses. Post-quorum, unweighted tallies drive local per-peer EWMA trust updates; cached records falling below \texttt{REPUTATION\_MIN\_TRUST} are subsequently excluded from local lookups.

\begin{algorithm}
\caption{Byzantine Voting (resolver cold path)}
\label{alg:voting}
\begin{algorithmic}
\STATE \textbf{Input:} domain $d$, reachable peer count $n$, timeout $T$, commit delay $\delta$
\STATE \textbf{Output:} IP address $ip$ or NULL
\STATE $h \leftarrow$ BLAKE2b-KEYED($d$, NETWORK\_PSK)
\STATE $q \leftarrow \max(\lfloor n/2 \rfloor + 1,\; \texttt{MIN\_QUORUM})$ \hfill \textit{(\texttt{MIN\_QUORUM}${=}3$)}
\STATE Broadcast signed VOTE\_REQUEST($h$)
\STATE $votes \leftarrow \{\}$; $seen\_keys \leftarrow \{\}$
\STATE $t_0 \leftarrow$ current\_time; $t_c \leftarrow t_0 + \delta$
\STATE \textbf{Phase 1 (collect):} buffer votes; defer quorum counting until $t_c$
\WHILE{current\_time $< t_c$ \textbf{and} current\_time $- t_0 < T$}
    \IF{signed VOTE\_RESPONSE for $h$ from public key $k$}
        \IF{$k \notin seen\_keys$ \textbf{and} verify\_signature \textbf{and} not equivocating($k,h$)}
            \STATE $seen\_keys \leftarrow seen\_keys \cup \{k\}$
            \STATE record\_binding($k, h$, response.ip)
            \STATE $votes[\text{response.ip}] \leftarrow votes[\text{response.ip}] + 1$
        \ENDIF
    \ENDIF
\ENDWHILE
\STATE \textbf{Phase 2 (commit):} continue collecting; evaluate identical-answer quorum
\WHILE{current\_time $- t_0 < T$}
    \STATE accept additional responses with the same rules as Phase~1
    \STATE $ip_{\max} \leftarrow \arg\max_{ip} votes[ip]$
    \IF{$votes[ip_{\max}] \geq q$}
        \STATE Update peer EWMA trust (reward agreeing keys, penalize divergent keys)
        \STATE cache.insert verified row for $d$ with $ip_{\max}$; \RETURN $ip_{\max}$
    \ENDIF
\ENDWHILE
\STATE Final quorum check on buffered votes; \RETURN NULL if $votes[ip_{\max}] < q$
\end{algorithmic}
\end{algorithm}

\subsection{Gossip-Based Cache Summary Announcements}

MeshDNS strictly separates authoritative resolution from cache awareness. While signed voting (\S\ref{sec:voting}) handles cold-cache misses, gossip implements a periodic \texttt{CACHE\_ANNOUNCE} channel (default 60~s). Announcements carry up to ten bounded summaries (keyed domain tag, IP, TTL, and reputation). When a node receives a high-reputation summary for an unknown tag from a peer exceeding \texttt{TRUST\_MIN\_ADMIT}, it issues a nonce-bearing unicast \texttt{CACHE\_REQUEST}. 

The peer replies with a signed \texttt{MSG\_CACHE\_RESPONSE} binding the domain tag, IP, TTL, reputation, requester nonce, and sender public key. The receiver accepts this prefetch row only if the Ed25519 signature verifies, the sender's key matches the PSK-admitted peer table, and the nonce matches an outstanding request. Crucially, these rows are marked non-authoritative: they cannot trigger resolver fast-path hits or produce signed \texttt{VOTE\_RESPONSE}s. This strict plane isolation ensures a poisoned prefetch cannot be amplified into honest votes; sustained gossip abuse degrades only cache-awareness utility, preserving the correctness of the voting plane.

\subsection{Threat Model}
\label{sec:threat}

We consider a Byzantine threat model where up to $f$ nodes in an $n$-node network may be compromised by adversaries. Compromised nodes may exhibit arbitrary malicious behavior, including providing \textit{False DNS responses} (returning incorrect IP addresses to redirect traffic), initiating \textit{Malicious vote responses} (attempting to poison peer caches via forged quorum answers), executing \textit{Vote manipulation} (colluding to influence quorum outcomes), and launching \textit{Denial of service} attacks (flooding the network with invalid packets). We assume an authenticated single-broadcast-domain deployment in which peers are admitted through a network pre-shared key. We also assume underlying cryptographic primitives (Ed25519, BLAKE2b) are secure. Under these assumptions, MeshDNS provides Byzantine fault isolation via authenticated per-query quorum resolution, rather than global Lamport-style Byzantine consensus.

\noindent\textbf{Scope of Byzantine Guarantees:} We explicitly bound our claims to \emph{Byzantine fault isolation} via authenticated majority agreement. MeshDNS's fault tolerance guarantees apply exclusively to \emph{intra-mesh} resolution among admitted peers: all peer votes are signature-verified, de-duplicated by peer identity, checked for equivocation, and subject to an identical-answer quorum before a result is accepted. We explicitly exclude the upstream recursive fallback path from our Byzantine threat model. When a distributed cache miss occurs, the node falls back to a configured upstream recursive resolver, inheriting the trust assumptions of that external infrastructure. Because the ESP8266 prototype cannot implement full local DNSSEC chain validation, a compromised upstream resolver could inject a poisoned record into a single node's cache on first lookup. 

To reduce this residual risk without the memory cost of on-device DNSSEC, the fallback path optionally cross-checks plain-DNS answers against Cloudflare DoH using the resolver-provided \texttt{AD} (Authenticated Data) flag. We note this cross-check is an architectural specification not implemented on the target ESP8266 hardware due to RAM limits, and its integration on higher-tier gateways is deferred to Section~\ref{sec:future}. When utilized, fallback results are cached only on a match or authenticated DoH agreement. Crucially, because MeshDNS enforces multi-peer identical-answer quorums, localized upstream poisoning is isolated and cannot autonomously propagate to infect the broader mesh.

\subsection{Security Mechanisms}

To defend against Byzantine attacks, MeshDNS employs a multi-layered security architecture: peer admission uses a network pre-shared key, and the signed control plane (\texttt{PING}/\texttt{PONG} discovery plus \texttt{VOTE\_REQUEST}/\texttt{VOTE\_RESPONSE}) uses Ed25519 signatures to prevent impersonation. At the core of the resolution process, signed quorum voting requires identical answers from a majority-derived quorum with a minimum floor of three peers. This voting path is hardened with duplicate-vote rejection, equivocation tracking, and a short commit delay. 

The 200~ms \texttt{VOTE\_COMMIT\_DELAY\_MS} provides a robust temporal buffer against late-vote timing manipulation. Because local subnet UDP round-trip times rarely exceed 20~ms, an adversary attempting to bypass Phase~1 equivocation tracking by timing a packet to arrive precisely at the commit boundary faces high stochastic network jitter. Packets arriving marginally early trigger standard equivocation defenses, while those arriving late are safely ignored, rendering targeted sub-millisecond temporal exploits practically infeasible on wireless IoT subnets. The system also respects standard DNS Time-to-Live (TTL) values, limiting the persistence of stale routing data.

\noindent\textbf{Trust and reputation.} Peer EWMA trust is adjusted after vote-processing events (successful quorum agreement, invalid signatures, equivocation, and duplicate keys) and gates gossip prefetch requests; quorum tallies remain unweighted. Per-entry cache reputation independently gates local cache hits and prefetch acceptance.

\subsection{Node Identity and Sybil Resistance}
A core vulnerability in decentralized local networks is the Sybil attack, where a single compromised host spoofs multiple network identifiers (e.g., MAC or IP addresses) to dominate a voting quorum. MeshDNS explicitly decouples node identity from easily spoofed network-layer headers. 

Instead, MeshDNS enforces identity at the cryptographic layer:
\begin{itemize}
    \item \textbf{Cryptographic Identity:} Upon initialization, each ESP8266 node generates and stores a persistent Ed25519 keypair in EEPROM. Quorum counting is strictly based on the number of valid, distinct Ed25519 signatures, ignoring MAC or IP duplication.
    \item \textbf{Admission Control:} Network participation is gated by a deployment-wide \texttt{NETWORK\_PSK}. This secret is used to authenticate discovery packets and key the BLAKE2b domain hashes within the voting protocol. 
    \item \textbf{Vote Integrity:} The protocol enforces strict ballot-stuffing protections (one vote per public key per resolution session) and detects equivocation (a single public key signing conflicting records), penalizing the offending node via a local trust matrix.
\end{itemize}

\noindent\textbf{Out-of-Scope Assumptions:} We assume a trusted provisioning phase where authorized physical nodes are flashed with the \texttt{NETWORK\_PSK}. Furthermore, because ESP8266-class devices lack secure hardware enclaves or TrustZone architectures, we assume the devices are physically secured within the deployment environment. Physical adversaries capable of executing JTAG memory dumps or EEPROM extraction to steal the plaintext \texttt{NETWORK\_PSK} fall outside our current local threat model. While our architecture prevents unauthenticated network-layer adversaries from injecting forged votes, mitigating physical insider Sybil attacks requires hardware-bound identity extraction (e.g., TPMs) as discussed in section \ref{sec:future}.

\subsection{Network Scalability Considerations}
Simultaneous cold-cache recovery on a shared Wi-Fi medium naturally contends for airtime. MeshDNS mitigates broadcast storms primarily through its strict voting timeouts and localized request boundaries. As demonstrated in Section~\ref{sec:evaluation}, the system degrades gracefully under application-level stress, trading successful query rates for stable latency when physical buffer exhaustion occurs.

\section{Implementation}
\label{sec:implementation}

\subsection{Hardware Platform and Firmware}

We implemented MeshDNS on the ESP8266 microcontroller, widely deployed in IoT due to its low cost and integrated Wi-Fi. It features an 80~MHz Tensilica L106 CPU and leaves 50~KB usable RAM after network initialization~\cite{espressif_esp8266ex_2023}. Operating at 80~mA active and 20~$\mu$A deep sleep, this platform accurately represents severe edge resource constraints, serving as an ideal evaluation baseline.

To facilitate empirical power profiling of the cryptographic and networking routines, the hardware testbed is integrated with an INA219 high-side current sensor \cite{ina219_datasheet} logging at 100~Hz. While dedicated high-frequency power profilers offer superior transient resolution, the INA219 was selected due to its widespread local-market availability and low cost, representing accessible hardware for edge deployments. To mitigate its sampling rate limitations when measuring sub-millisecond cryptographic operations, we employed a loop-aggregation microbenchmarking technique. The $\sim$3500-line firmware is structured into modular controllers with aggressive static allocation to remain within the 50~KB usable RAM envelope (Appendix~\ref{app:implementation}).

\subsection{Network Protocol}

MeshDNS uses UDP communication within a single local subnet, following the deployment model of mDNS-style local IoT networks while avoiding continuous multicast group maintenance on the ESP8266. In practice, the implementation uses local subnet broadcast for discovery and voting paths, plus unicast command/telemetry flows for repeatable hardware experiments.

Messages use compact variable-length binary encoding (domain tag, addresses, timestamps, and 64-byte Ed25519 signatures) and remain within Wi-Fi MTU limits. Core types include \texttt{VOTE\_REQUEST} (0x05), \texttt{VOTE\_RESPONSE} (0x06), \texttt{CACHE\_ANNOUNCE} (0x03), and \texttt{CACHE\_REQUEST} (0x04); peer admission uses PSK-authenticated \texttt{PING}/\texttt{PONG} discovery.

\subsection{Cryptographic Implementation}
To accommodate the computational limits of the ESP8266, the prototype utilizes lightweight yet standard cryptographic primitives. The specific implementations are as follows:

\smallskip
\noindent\textbf{Hashing:} Domain names are converted into keyed \textbf{BLAKE2b} tags~\cite{rfc7693} using the mesh \texttt{NETWORK\_PSK}. This preserves the compact 32-byte representation while reducing offline dictionary enumeration by observers that do not possess the mesh key.

\smallskip
\noindent\textbf{Signatures:} The signed control plane---peer discovery (\texttt{PING}/\texttt{PONG}), cooperative voting (\texttt{VOTE\_REQUEST}/\texttt{VOTE\_RESPONSE}), and unicast gossip prefetch replies (\texttt{CACHE\_RESPONSE})---uses \textbf{Ed25519} signatures, leveraging their compact 64-byte size and efficient Curve25519-based verification on the ESP8266's 80~MHz core. \texttt{CACHE\_RESPONSE} signatures are nonce-bound to the corresponding \texttt{CACHE\_REQUEST} to prevent replay. Firmware statistics report accumulated sign/verify/hash microseconds per operation (\texttt{TRACK\_CRYPTO\_COST}).

\smallskip
\noindent\textbf{Random Number Generation (RNG):} PSK-authenticated \texttt{PING}/\texttt{PONG} discovery nonces are drawn from the hardware-backed \texttt{os\_random()} API and tracked in an outstanding-nonce table for replay resistance. Signed vote packets use \texttt{millis()} timestamps in the signed payload (freshness within the voting timeout), not random query IDs.

\smallskip
\noindent\textbf{Authenticated Encryption:} \textbf{XChaCha20-Poly1305}~\cite{rfc8439} is available through the Monocypher wrapper for authenticated encryption support in higher-security deployment modes.

\section{Evaluation}
\label{sec:evaluation}

We evaluate MeshDNS using a hardware prototype and complementary discrete-event simulation. The hardware experiments quantify local resolution latency, signed-quorum overhead, adversarial quorum behavior, and stress performance on ESP8266 devices. The simulation experiments extend the analysis to larger populations and churn scenarios that exceed the scale of the physical testbed.

\begin{figure*}[t]
    \centering
    \begin{subfigure}[b]{0.28\linewidth}
        \centering
        \includegraphics[width=\linewidth]{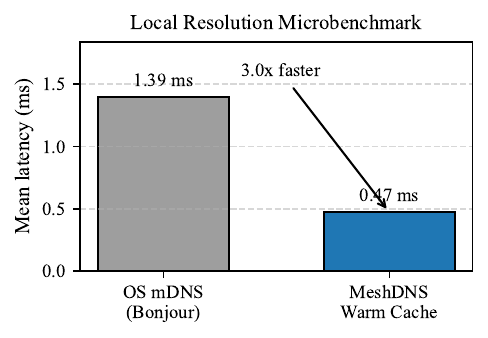}
        \caption{OS mDNS baseline vs. MeshDNS.}
        \label{fig:microbenchmark}
    \end{subfigure}
    \hfill 
    \begin{subfigure}[b]{0.38\linewidth}
        \centering
        \includegraphics[width=\linewidth]{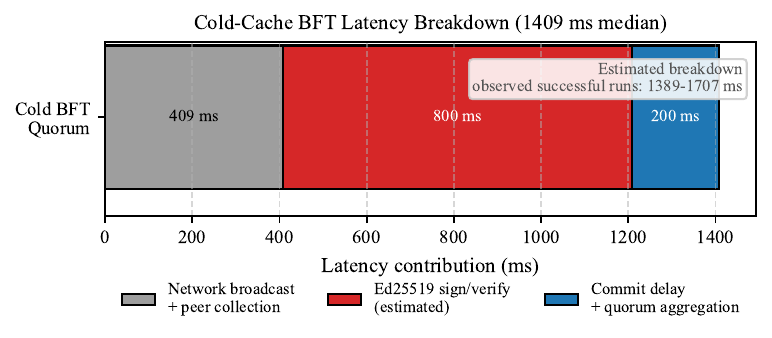}
        \caption{Latency breakdown of cold-quorum.}
        \label{fig:crypto_penalty}
    \end{subfigure}
    \hfill 
    \begin{subfigure}[b]{0.32\linewidth}
        \centering
        \includegraphics[width=\linewidth]{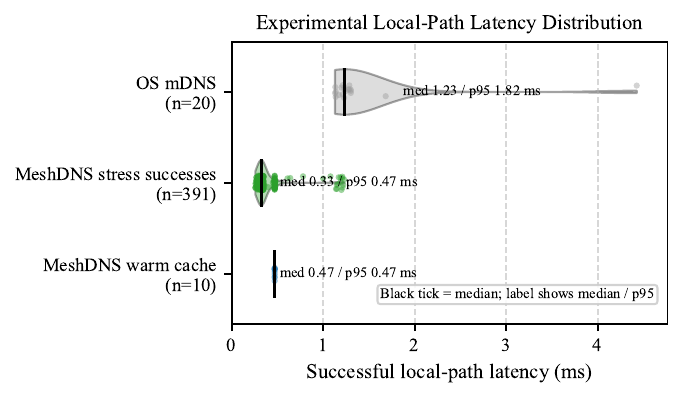}
        \caption{CDF of DNS resolution latency.}
        \label{fig:latency_cdf}
    \end{subfigure}
    
    \caption{Local resolution performance on the ESP8266 testbed. (a) A direct comparison demonstrating the sub-millisecond efficiency of the warm-cache fast path. (b) Execution timeline of the cold-quorum penalty. Ed25519 time is labeled ``Estimated'' because while the firmware precisely tracks raw cryptographic CPU cycles, the visual timeline aggregates these with non-deterministic RTOS and LwIP context switching. (c) Distribution of resolution latencies, highlighting the overhead of cold signed-quorum resolution.}
    \label{fig:latency_panel}
\end{figure*}

\subsection{Experimental Setup and Methodology}

To comprehensively evaluate MeshDNS, we employed a hybrid methodology combining empirical measurements from a physical ESP8266 deployment with discrete-event simulations for larger topologies.

\textbf{Hardware Testbed:} We deployed five ESP8266 NodeMCU microcontrollers on a single 802.11n broadcast domain. The workload exercises four resolution modes: (1) a local mDNS baseline, (2) warm MeshDNS cache hits, (3) cold signed-quorum resolution after clearing the resolver cache, and (4) a burst-style stress workload against the warm-cache path. The canonical target was an mDNS-style \texttt{.local} hostname on a fixed LAN address. For benchmark names, firmware disables upstream DNS fallback so failed peer quorums are explicitly reported as \texttt{peer\_failed} rather than masked by root DNS. Host-side scripts automate cache clearing, resolution, and telemetry over UDP ports 8080--8081 for data collection, completely separated from peer protocol traffic on port 5353. Adversarial behavior is injected by reflashing individual nodes with \texttt{BYZANTINE\_MODE} and level-specific attack parameters.

\textbf{Testbed scale:} We evaluate on $N{=}5$ nodes, matching typical smart-home IoT cells (which often feature three to six active edge devices) and stressing the prototype's minimum quorum floor (\texttt{MIN\_QUORUM}${=}3$). Because the protocol enforces a strict identical-answer acceptance rule, adding further honest hardware nodes would merely generate redundant identical votes without exercising any new cryptographic or Byzantine logic. Consequently, expanding the physical hardware array significantly inflates reflashing and management overhead without yielding additional security insights.

\textbf{Discrete-Event Simulation:} We utilize a Python \texttt{SimPy} model to evaluate cache-summary propagation, churn, and coverage at $N \in \{50, 250, 1000\}$. The model uses five random seeds per point and incorporates median hardware calibrations (0.47~ms warm cache, 1.41~s cold quorum, and a 0.218 UDP drop rate derived from stress telemetry). Default configurations inject 15\% Byzantine nodes and 20\% network churn per interval, with no simulated root fallback. The simulator abstracts RF interference and 802.11 MAC behavior, serving strictly as a \emph{protocol-level stress bound} on churn and coverage, rather than a claim that a single 802.11 broadcast domain routinely hosts thousand-node deployments.

\subsection{Microbenchmarks and Resolution Latency}

Figure~\ref{fig:latency_cdf} presents the latency distribution observed in the hardware experiments. The results separate warm-cache resolution from cold signed-quorum resolution because these modes exercise different parts of the system: warm-cache hits measure the local fast path, whereas cold quorum resolution measures authenticated peer agreement after a local cache miss.

Warm MeshDNS cache hits average 0.47~ms, compared with 1.39~ms for the local mDNS baseline (Figure~\ref{fig:microbenchmark}). This demonstrates that the cache-hit path adds negligible overhead once a record is already present locally. By contrast, cold signed-quorum resolution completes in 1300-1500 ms (median 1409~ms) with a 3/3 quorum. This higher latency reflects peer voting, signature verification, and timeout handling, and should be interpreted as the upfront cost of authenticated peer agreement rather than as a replacement for the warm-cache fast path.

Figure~\ref{fig:crypto_penalty} breaks down the median 1409~ms cold path; Ed25519 sign/verify dominates on the 80~MHz core, consistent with firmware \texttt{TRACK\_CRYPTO\_COST} counters. The warm-cache fast path (0.47~ms) amortizes this cost for repeated local lookups. Crucially, this is a strict, one-time upfront cost for authenticated peer agreement; once a quorum is validated, the record is cached locally, rendering all subsequent lookups virtually instantaneous via the warm-cache fast path.

The $\sim$1.3--1.7\,s cold-quorum penalty represents a strict \emph{security--latency trade-off}. Its impact heavily depends on the workload. For \textbf{steady-state telemetry} (e.g., an MQTT sensor publishing every five minutes with a 24-hour TTL), the amortized cost drops to $\sim$5.6~ms per query, highly favoring MeshDNS. Conversely, \textbf{boot-time recovery} or \textbf{interactive discovery} workloads pay the full penalty per unresolved name. For these latency-sensitive modes, we do not claim sub-second cold resolution; applications requiring instant discovery must either maintain their own persistent state across reboots or simply accept the upfront delay as the necessary cost of authenticated peer agreement in untrusted environments.

\subsection{System Resilience Under Stress}

We next evaluate the system's behavior and communication footprint under a burst of warm-cache resolution requests. In the hardware stress experiment, the resolver issued 500 requests at an injection rate of 11.08 requests/s. As summarized in Table~\ref{tab:stress}, the system processed 391 successful cache-hit responses, producing a 78.2\% cumulative hit yield. 

\begin{table}[htbp]
\centering
\scriptsize 
\setlength{\tabcolsep}{3pt} 
\caption{Stress Workload and Communication Footprint}
\label{tab:stress}
\begin{tabular}{@{}l c l c@{}}
\toprule
\textbf{Metric} & \textbf{Value} & \textbf{Metric} & \textbf{Value} \\ \midrule
Req. sent & 500 & Mean latency & 0.37 ms \\
Succ. responses & 391 & Cmd traffic & 13.82 kbps \\
Success rate & 78.2\% & Telemetry traffic & 14.27 kbps \\
Injection rate & 11.08 req/s & Heartbeat traffic & 0.56 kbps \\ \bottomrule
\end{tabular}
\end{table}

\begin{figure}[htbp]
    \centering
    \begin{subfigure}[b]{0.48\linewidth}
        \centering
        \includegraphics[width=\linewidth]{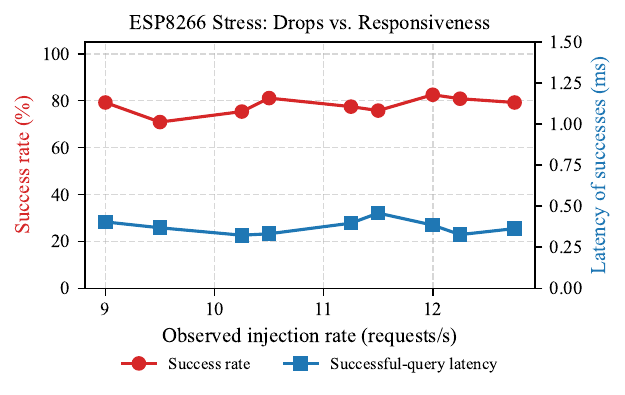}
        \caption{Success rate vs latency.}
        \label{fig:stress_dual_axis}
    \end{subfigure}
    \hfill 
    \begin{subfigure}[b]{0.48\linewidth}
        \centering
        \includegraphics[width=\linewidth]{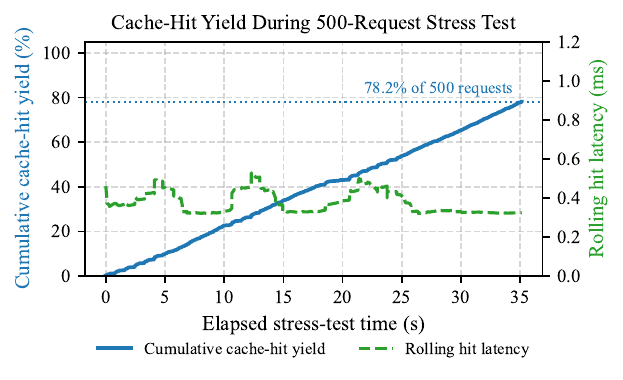}
        \caption{Cumulative hit yield.}
        \label{fig:cache_hit}
    \end{subfigure}
    
    \caption{System Resilience Under Hardware Stress. During a 500-request burst, the ESP8266 testbed degrades gracefully under LwIP buffer exhaustion. (a) Successful-query latency remains completely stable despite dropping packets. (b) The system steadily yields a 78.2\% overall success rate.}
    \label{fig:stress_panel}
\end{figure}

As shown in Figure~\ref{fig:stress_dual_axis}, while the success rate degrades under heavy burst pressure due strictly to LwIP buffer socket exhaustion (not cache TTL expiry) on the constrained ESP8266, the system degrades gracefully. The latency of the successful responses remains completely stable, averaging 0.37~ms, and stabilizing near 0.32~ms toward the end of the run (Figure~\ref{fig:cache_hit}). 

Furthermore, the combined command, telemetry, and heartbeat traffic generated during this workload remains exceptionally low, drawing under 30~kbps of total network bandwidth. This indicates that MeshDNS control traffic remains modest at the application level, providing an efficient footprint for IoT deployments without triggering broadcast storms. Note that this result reflects a warm-cache stress measurement and bounding of application-level control traffic; gossip traffic consists of bounded summaries and occasional unicast prefetch exchanges, separate from on-demand voting packets.

\subsection{Byzantine Fault Isolation}

To evaluate resilience against malicious internal actors, we use a focused hardware run. In a baseline honest scenario, cold quorum resolution succeeds on the first attempt (median 1409~ms). 

To model an adversarial environment, one admitted peer was configured to return a fixed incorrect IP (\texttt{6.6.6.6}). Because the resolver strictly enforces an identical-answer quorum, it successfully detected and rejected the conflicting adversarial vote. This forced quorum aggregation to fail, triggering two consecutive timeouts before successfully securing a 3/3 honest quorum on the third attempt. 

While this demonstrates reliable fault isolation and cache poisoning prevention, it highlights the severe temporal retry penalty inherent to adversarial environments. Crucially, because MeshDNS enforces PSK-based admission and Ed25519 signatures, this accurately models a compromised, admitted peer rather than an unconstrained Sybil attacker. Section~\ref{sec:adversarial-eval} extends this analysis with a structured attack matrix.

\subsection{Systematic Adversarial Evaluation}
\label{sec:adversarial-eval}

To complement the focused isolation experiment above, we executed a structured adversarial matrix on the five-node ESP8266 testbed and a calibrated \texttt{SimPy} model reusing the same cold-quorum timing distributions. The test script accepts a configurable round count; all curated spot-checks reported here utilize five cold-resolution rounds per scenario. In each round, caches are cleared on honest voters, records are seeded with the correct LAN address, and the designated resolver issues \texttt{CMD\_RESOLVE} while host-side telemetry records success, timeout, false acceptance (incorrect IP delivery), and signed-quorum latency. Byzantine nodes are configured in firmware; the resolver remains honest, and at least three distinct honest voters participate except where the topology cannot supply them.

\textbf{Hardware spot-checks.} Table~\ref{tab:adversarial_hw} summarizes curated runs on the deployment. Reported success rates are availability metrics; false-accept counts represent the primary security readout.

\begin{table}[htbp]
\centering
\caption{Curated hardware adversarial spot-checks ($N{=}5$ ESP8266 nodes). Success and false-accept columns report $k/5$ cold trials.}
\label{tab:adversarial_hw}
\resizebox{\columnwidth}{!}{%
\begin{tabular}{@{}l l c c c@{}}
\toprule
\textbf{Scenario} & \textbf{Attack} & \textbf{Succ.} & \textbf{False acc.} & \textbf{Med.\ $T_q$ (ms)} \\ \midrule
$f{=}0$ benign & none & 3/5 & 0/5 & 1489 \\
$f{=}1$ L1 & \texttt{6.6.6.6} & 4/5 & 0/5 & 1645 \\
$f{=}2$ L1 & \texttt{6.6.6.6} & 0/5 & 0/5 & --- \\
L4 equivoc. & dual wrong IP & 3/5 & 0/5 & 1847 \\
Sybil $k{=}3$ & multi-key & 4/5 & 2/5 & 1612 \\ \bottomrule
\end{tabular}%
}
\end{table}

Under benign operation ($f{=}0$), cold quorum succeeded in 3/5 rounds with zero false accepts; the remaining rounds timed out without accepting an incorrect address. With one Level-1 Byzantine node returning a fixed poisoned address, success reached 80\% (4/5) while false acceptance remained strictly at 0/5. These hardware spot-checks validate the firmware execution path but should be interpreted as small-sample availability measurements under RF and harness effects rather than statistical convergence experiments.

The integrity result aligns with the underlying architecture: identical-answer quorum enforcement relies on deterministic cryptographic state machines (Ed25519 verification), not stochastic heuristics. The mathematical rejection of conflicting signatures produces the same state transitions across trials, while the observed success rate is still shaped by network-layer stochasticity such as packet loss and timeout scheduling. We constrain that remaining stochasticity via the discrete-event simulation sweeps below.

At $f{=}2$ on $N{=}5$, the mathematical impossibility of forming a 3/3 honest quorum correctly resulted in 100\% \texttt{peer\_failed} outcomes (0/5 success). As expected from quorum arithmetic, this represents an availability limit, not a Byzantine isolation failure. Level-4 equivocation tests rejected conflicting payloads without false acceptance, yielding 60\% success (3/5) and 40\% safe timeouts.

\textbf{Sybil identities.} We stress-tested one admitted node presenting three independent, synthetic Ed25519 keypairs (authorized via the mesh \texttt{NETWORK\_PSK} admission gate). Across the curated five-round Sybil spot-check, quorum resolution completed in 4/5 rounds, but 2/5 rounds accepted \texttt{6.6.6.6}. This negative result shows that PSK-based admission and per-key de-duplication alone are insufficient when one physical node can mint multiple admitted identities, reiterating the need for gateway-enforced identity caps in massive deployments (Section~\ref{sec:threat}).

\textbf{Simulation sweep.} Another discrete-event test script replays Level-1 attacks for $N \in \{5, 7, 10\}$. Under zero packet loss, quorum success remains 100\% with 0\% false acceptance across all valid $(N, f)$ pairs, isolating protocol logic from RF effects. Collusion and packet-loss suites similarly maintain zero false acceptance when honest peers hold correct records.

\subsection{Protocol-Level Churn Extrapolation}

\begin{figure}[htbp]
\centering
\includegraphics[width=\linewidth]{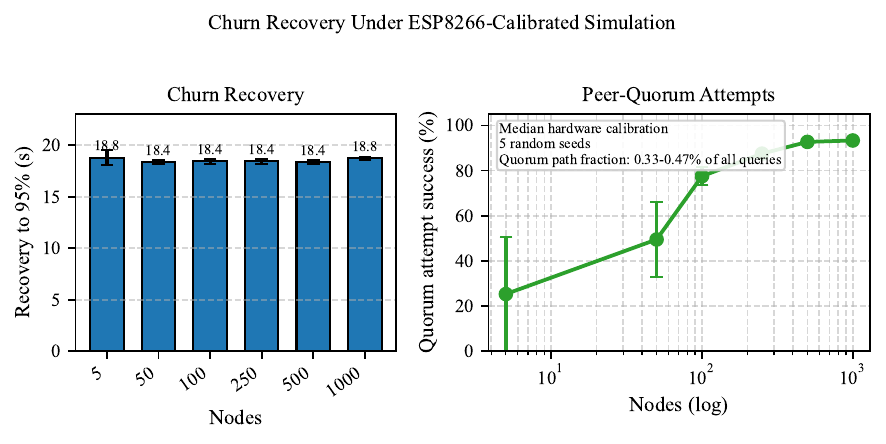}
\caption{Churn recovery and peer-quorum attempt success versus population size in the calibrated \texttt{SimPy} model (five seeds per $N$; 95\% confidence intervals). \textbf{Left:} time to recover 95\% cache coverage after churn. \textbf{Right:} fraction of cold-resolution attempts that achieve a valid signed quorum; the quorum path remains a small share (${\sim}0.33$--$0.47\%$) of all queries.}
\label{fig:scalability}
\end{figure}

To evaluate MeshDNS beyond the physical testbed, we utilize a \texttt{SimPy} discrete-event model with \textit{median} hardware calibration (five random seeds per $N$) to study churn recovery and quorum behavior from $N{=}5$ to $N{=}1000$ nodes. The upper population sizes represent an extreme scalability stress test; real single-broadcast-domain IoT cells are far smaller. We explicitly note these 1,000-node results represent theoretical protocol-level extrapolation bounds, not literal deployment targets, as 802.11 MAC-layer beacon congestion would physically dominate a single broadcast domain at this scale. Figure~\ref{fig:scalability} visualizes the results.

Figure~\ref{fig:scalability} shows recovery to 95\% coverage remains stable near 18.4--18.9~s across population sizes, indicating churn re-convergence does not grow materially with $N$. Peer-quorum \textit{attempt} success rises from ${\sim}13\%$ at $N{=}5$ to ${\sim}93\%$ at $N{\geq}500$, as larger meshes provide more reachable voters even though only ${\sim}0.46\%$ of queries enter the cold path. The $N{=}5$ row exposes a 60\% coverage ceiling---below the three-peer BFT floor---highlighting small populations as a stress case rather than a production target. Because the simulator abstracts RF interference, we treat these results strictly as a protocol-level churn study rather than a literal Wi-Fi forecast. Ultimately, the simulation supports our localized broadcast-domain design intent: MeshDNS provides predictable churn recovery for compact IoT cells, with massive-scale federation deferred to future hierarchical gateway designs.

\subsection{Comparative Analysis with Existing Protocols}

To contextualize MeshDNS within the broader landscape of naming systems, Table~\ref{tab:protocol_comparison} compares MeshDNS against traditional and decentralized protocols. MeshDNS values represent prototype hardware measurements, while non-MeshDNS values are aggregated from established literature context.

\begin{table*}[t]
\centering
\caption{Protocol comparison (N/A* not feasible on ESP8266). MeshDNS rows are measured on our 5-node testbed; other protocols use literature ranges on different topologies and hardware---not head-to-head benchmarks.}
\label{tab:protocol_comparison}
\begin{tabular}{@{}lcccccc@{}}
\toprule
\textbf{Protocol} & \textbf{Latency (LAN)} & \textbf{Energy/Query} & \textbf{Fault Tolerance} & \textbf{Memory} & \textbf{Scope} & \textbf{Decentralized} \\ \midrule
Traditional DNS & 80-120 ms & not evaluated & Centralized SPOF & low client state & Internet & No \\
mDNS & 1.39 ms measured & not evaluated & None & low client state & Local LAN & Partial \\
Namecoin~\cite{kalodner2015namecoin} & 10--60 min & N/A* & 51\% attack & 10+ GB & Internet & Yes \\
Ethereum ENS~\cite{wood2014ethereum} & 12--15 sec & N/A* & 51\% attack & 40+ GB & Internet & Yes \\
Standard DHT~\cite{falkner2007profiling, trautwein2022ipfs} & 80--150 ms & N/A* & Churn resilient & MB-scale & Wide-area overlay & Yes \\
\textbf{MeshDNS warm} & \textbf{0.47 ms} & \textbf{2~$\mu$J} & local cache & 36 KB & single LAN & Yes \\
\textbf{MeshDNS cold quorum} & \textbf{1300-1500 ms} & \textbf{active radio TX} & signed 3/3 quorum & 36 KB & single LAN & Yes \\ \bottomrule
\end{tabular}
\end{table*}

As demonstrated by Table~\ref{tab:protocol_comparison}, MeshDNS uniquely occupies a favorable region of the latency--memory--fault-tolerance tradeoff space for resource-constrained \emph{single-LAN} deployments. Cross-protocol latency numbers are indicative only: direct comparison would require implementing each system on the same ESP8266 testbed under identical RF conditions. Blockchain-based systems (Namecoin, ENS) impose prohibitive confirmation latencies (10--60~s) and massive storage footprints (40+~GB) that are entirely incompatible with IoT hardware. Similarly, DHT-based approaches require megabytes of routing state, sitting two orders of magnitude beyond the ESP8266's memory capacity. Finally, while mDNS fits the memory constraints, it provides neither Byzantine-resilient signed quorums nor internet-wide routing capabilities. MeshDNS successfully bridges this gap, delivering sub-millisecond warm-cache latency, a minimal 36~KB memory footprint, and signed, Byzantine-resilient local quorum behavior natively tailored for a single broadcast domain.

\subsection{Energy Consumption}

\begin{figure}[htbp]
\centering
\includegraphics[width=0.85\linewidth]{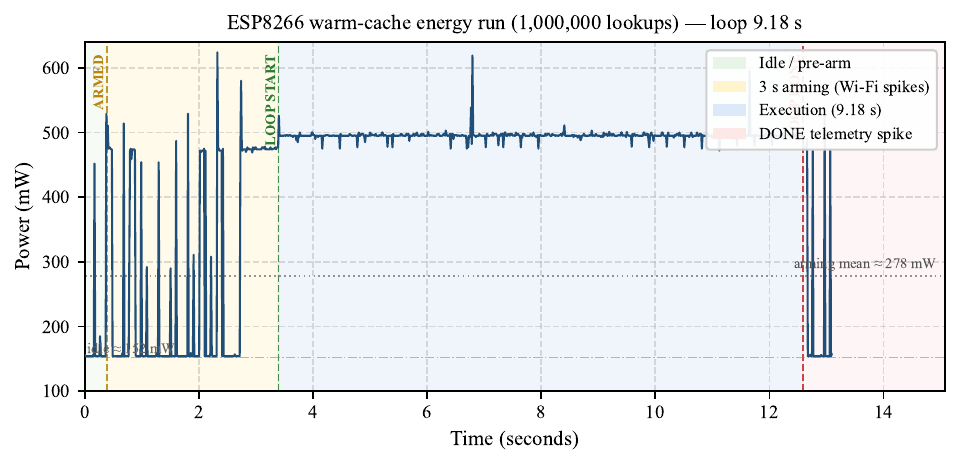}
\caption{Empirical power consumption trace of the ESP8266 node during a 1,000,000-lookup loop-aggregation microbenchmark. The trace delineates the 152~mW network idle baseline, the active Wi-Fi arming spikes, and the sustained CPU execution plateau during pure cache resolution.}
\label{fig:power_trace}
\end{figure}

IoT energy consumption is dominated by Wi-Fi radio activation (TX/RX). Baseline profiling (Figure~\ref{fig:power_trace}) confirms our ESP8266 draws ${\sim}$152~mW while network-idle, spiking to ${\sim}$500~mW during UDP broadcasts. MeshDNS bypasses these transmissions via cooperative caching. To isolate pure cache-lookup CPU overhead from network latency, we executed a 1,000,000-iteration internal microbenchmark in 9.17~s. Subtracting the integrated idle power from the gross loop energy reveals that a raw SRAM memory-path lookup consumes just 2~$\mu$J.

By isolating transmission-intensive quorum resolution strictly to the cold path, MeshDNS makes localized BFT energetically viable for edge deployments. A modeled cold-path quorum (50~ms TX at ${\sim}$500~mW plus a 3.5~s receive window at 152~mW) consumes a lower bound of 557~mJ, scaling linearly if adversarial timeouts force retry loops. Compared to the 2~$\mu$J warm-cache hit, assuming an illustrative 80\% cache-hit rate, the amortized cost drops to ${\sim}$111~mJ per query.
Finally, cryptographic profiling of the hardened gossip plane quantifies the signed-prefetch overhead. The bounded nonce table and 142-byte \texttt{CACHE\_RESPONSE} buffers consume 832 bytes of static heap RAM. A single Ed25519 sign-and-verify prefetch exchange requires 415~ms of CPU time; correlated with our INA219 telemetry (156~mW active compute draw), this consumes 64.7~mJ. Executed asynchronously every 60 seconds, this secures the gossip awareness channel without inflating the 0.47~ms warm-cache fast path.

\section{Discussion}
\label{sec:discussion}

\subsection{Security and Threat Resilience}

The security contribution of MeshDNS is best understood as Byzantine-resilient
local resolution rather than full Byzantine consensus. By scoping agreement to
individual DNS lookups, the system avoids the prohibitive overhead of global
consensus protocols, making it viable for sub-50~KB RAM edge devices.

Architecturally, we demonstrate that identical-answer, Ed25519-authenticated
quorums---combined with PSK-keyed domain tags and duplicate-vote
rejection---successfully secure local resolution. Our empirical results validate
this threat boundary: the protocol strictly isolates compromised nodes during
Level-1 and Level-4 attacks. However, as the Sybil stress test highlights,
PSK-based admission alone cannot cap identities per device. True Sybil immunity
in massive deployments requires stronger hardware-bound admission (e.g., TPMs),
though our lightweight approach remains highly effective for standard, trusted
IoT cells.

Crucially, these guarantees protect \emph{intra-mesh} resolution. MeshDNS does
not inherently validate records initially obtained from untrusted upstream
resolvers, making the fallback path a separate trust boundary. The fallback path
is designed to optionally cross-check plain-DNS answers against DoH using the
resolver-provided \texttt{AD} flag, though this cross-check is not implemented
on the target ESP8266 hardware and is deferred to higher-tier gateway
integration (Section~\ref{sec:future}). Regardless, because MeshDNS enforces
multi-peer identical-answer quorums, any upstream-poisoned record injected into
a single node's cache is isolated and cannot propagate to infect the broader
mesh. Deployments requiring absolute first-lookup guarantees still necessitate
end-to-end local DNSSEC validation.

\noindent\textbf{Gossip-plane hardening.} Prefetch replies are signed and
nonce-bound, and the receiver checks that the signing key matches the
PSK-admitted peer identity for the source IP. The firmware also treats gossip
rows as non-authoritative: rows marked \texttt{gossip\_prefetch} are ignored
when answering \texttt{VOTE\_REQUEST} packets. This prevents an admitted peer
from poisoning hash-only prefetch state and causing honest nodes to amplify the
wrong binding as signed votes.

Finally, privacy is bounded by the local membership model. Keyed BLAKE2b tags
mitigate offline dictionary enumeration by non-members, offering an optimal
balance between operational privacy and hardware constraints. Cryptographically
heavy alternatives like differential privacy or private information retrieval
(PIR) remain infeasible for ESP8266-class hardware.

\subsection{Deployment and Applicability}

MeshDNS targets compact IoT deployments sharing a single 802.11 broadcast domain (e.g., smart homes or industrial LANs). This central design assumption allows nodes to discover peers, exchange signed votes on demand, and reuse records without reaching external resolvers; periodic gossip supplements authenticated cache awareness but does not substitute for the signed voting path. The system is therefore most applicable to settings where local survivability, low warm-cache latency, and authenticated peer agreement are more important than wide-area peer connectivity.

This scope also clarifies the relationship between MeshDNS and a conventional access-point-based DNS resolver. In benign networks where the access point is trusted, continuously available, and easy to manage, a local resolver at the AP may be simpler. However, MeshDNS targets a fundamentally different operating point: environments where network infrastructure is untrusted, transient, or locked down (e.g., ISP-provisioned hardware). By decoupling resolution state from the access point and distributing it across the edge devices themselves, nodes can cross-check cached answers and survive centralized infrastructure outages. In this sense, MeshDNS complements local DNS infrastructure rather than universally replacing it.

Architecturally, MeshDNS is strictly optimized for intra-subnet environments. Complex wide-area routing, NAT traversal, and eclipse-resistant cross-domain sampling are deliberately deferred to future hierarchical gateway designs. While our simulation results demonstrate robust protocol-level convergence under churn, translating this to massive multi-domain deployments will require explicit gateway clustering, hierarchical membership management, and a careful definition of inter-cluster trust relationships.

\section{Limitations and Future Work}
\label{sec:future}

While our prototype successfully validates the theoretical framework, several limitations inform our open-source roadmap and future research trajectory.

\subsection{Hardware Extensions and Simulation Fidelity}
The current prototype operates exclusively over IPv4 and lacks full local DNSSEC chain validation—the primary residual trust boundary identified in Section~\ref{sec:threat}. As MeshDNS transitions to an open-source framework, developers can extend support to edge devices with higher RAM and CPU capacities (e.g., ESP32-class hardware). This will enable true on-device DNSSEC validation, 128-bit IPv6 addressing, and production-grade cryptographic profiling. Furthermore, while our \texttt{SimPy} model provides a protocol-level stress bound, future evaluations should transition to high-fidelity network simulators (e.g., \texttt{ns-3}, OMNeT++) and large-scale physical testbeds to measure true RF coexistence and quorum jitter under mass recovery.

\subsection{Security Evolution and Optimizations}
To mitigate physical insider Sybil attacks, the framework will serve as a foundation for integrating hardware-bound identities (e.g., TPMs), while formal verification (e.g., via TLA+) will strengthen theoretical guarantees against complex Byzantine strategies. Architecturally, hierarchical gateway clustering will help transcend the congestion limits of a single 802.11 broadcast domain, facilitating integration with federated edge platforms (EdgeX Foundry~\cite{edgex_foundry}, KubeEdge~\cite{kubeedge}) and infrastructure-less networks (LoRaWAN, IEEE 802.11s). Finally, communication overhead under severe broadcast contention can be further reduced by implementing compressed Bloom filters, frequency-based prefetching, and MAC-layer backoff mechanisms.

\section{Conclusion}
\label{sec:conclusion}

This paper presented MeshDNS, a cooperative DNS resolution framework tailored for resource-constrained IoT networks. By integrating distributed caching, authenticated quorum voting, and hash-based cache awareness, MeshDNS successfully decouples local name resolution from centralized network infrastructure. Our ESP8266 firmware implementation demonstrates the practicality of deploying secure distributed protocols on commodity edge hardware constrained to sub-50~KB of usable RAM. Validated through a hybrid evaluation methodology, the system achieves a 0.47~ms warm-cache fast path---outperforming native mDNS baselines---while transparently trading a one-time $\sim$1.3--1.7\,s initialization penalty to successfully isolate Byzantine injections during cold-quorum resolution. Structured adversarial sweeps confirm strict protocol safety, reporting zero false acceptance across Level-1, Level-4(Section~\ref{sec:adversarial-eval}).

Ultimately, MeshDNS addresses the critical vulnerabilities of centralized DNS in edge environments, providing a highly resilient foundation for autonomous IoT infrastructure. We explicitly condition these fault-isolation guarantees on the physical security of the deployment: because ESP8266-class hardware stores the shared admission key in plaintext EEPROM, physical device extraction trivially yields full admission credentials and falls outside our threat model. MeshDNS's security properties therefore apply strictly to network-layer adversaries and compromised-but-admitted peers acting within the cryptographic protocol. 

To facilitate reproducibility and peer research, the complete ESP8266 firmware, SimPy simulation environments, and empirical datasets are open-source and publicly available at \href{https://github.com/mahbubasif/MeshDNS-Artifact}{https://github.com/mahbubasif/MeshDNS-Artifact}.

\appendices

\section{Resolver Cache Lookup Algorithm}
\label{app:algorithms}

Application-facing resolution uses plaintext domain-string lookup; gossip prefetch rows (synthetic \texttt{\#g:...} names) are excluded from this fast path and are non-authoritative hints. The firmware marks them as \texttt{gossip\_prefetch}, and \texttt{VOTE\_REQUEST} handling refuses to cast signed votes from those rows.

\begin{algorithm}
\caption{Resolver cache lookup (fast path)}
\label{alg:cache_lookup}
\begin{algorithmic}
\STATE \textbf{Input:} domain name $d$
\STATE \textbf{Output:} IP address $ip$ or NULL
\STATE $entry \leftarrow$ cache.find\_by\_domain($d$) \hfill \textit{(string key)}
\IF{$entry =$ NULL}
    \RETURN NULL
\ENDIF
\IF{$entry.reputation < $ REPUTATION\_MIN\_TRUST}
    \RETURN NULL
\ENDIF
\IF{$entry.timestamp + entry.ttl > $ current\_time}
    \RETURN $entry.ip$
\ELSE
    \STATE cache.remove($entry$)
    \RETURN NULL
\ENDIF
\end{algorithmic}
\end{algorithm}

\section{Firmware Architecture and Memory Profiling}
\label{app:implementation}

\subsection{Software Architecture}
The firmware prototype consists of approximately 3500 lines of C++ code. To accommodate the strict memory limitations of the ESP8266, the software is organized into tightly coupled logical modules:

\begin{itemize}
    \item \textbf{Core Controller:} Manages WiFi initialization, the primary DNS resolution loop, UDP command handling, and network statistics reporting.
    \item \textbf{Distributed Cache:} Implements the LRU cache using a hash table with linear probing and keyed domain tags to minimize memory fragmentation and reduce offline domain enumeration.
    \item \textbf{Signed Quorum Engine:} Operates the Byzantine-resilient voting manager, handling timeout-based response collection, duplicate-vote rejection, equivocation tracking, commit delay, and quorum validation.
    \item \textbf{Gossip Summary Manager:} Broadcasts \texttt{CACHE\_ANNOUNCE} summaries and, on high-trust hints, issues \texttt{CACHE\_REQUEST}/\texttt{CACHE\_RESPONSE} to populate signed, nonce-bound hash-indexed prefetch rows for cache awareness; rows marked \texttt{gossip\_prefetch} cannot cast votes, and signed \texttt{VOTE\_REQUEST}/\texttt{VOTE\_RESPONSE} remains the resolver's authoritative cold path.
    \item \textbf{Network and Admission Stack:} Manages UDP packet processing, local subnet broadcast/unicast communication, PSK-authenticated peer discovery, nonce replay resistance, telemetry, and binary message serialization.
    \item \textbf{Cryptographic Wrapper:} Provides Monocypher-backed implementations of Ed25519 signatures, keyed BLAKE2b tags, and XChaCha20-Poly1305 authenticated encryption.
\end{itemize}

\subsection{Memory Optimization}
The strict 50~KB usable RAM limit necessitates aggressive memory management. Key optimizations include: \textbf{(1) Static allocation} using pre-allocated arrays for all data structures to prevent heap fragmentation; \textbf{(2) bounded cache-summary announcements} rather than transmitting full domain lists; \textbf{(3) Compact encoding} via binary serialization, saving 60\% in message size compared to JSON; and \textbf{(4) Configurable cache sizing}, defaulting to 20 entries in firmware. Overall, the framework consumes approximately 36~KB of RAM (cache, network buffers, voting state, and gossip), leaving margin for the WiFi stack and dynamic overhead.

\section*{LLM Usage Statement}
LLMs were used strictly for editorial refinement and LaTeX formatting. All manuscript text, system architecture design, and firmware/simulation source code are the exclusive, original work of the authors. As required by the conference guidelines, all LLM outputs were inspected by the authors to ensure accuracy and originality.

\bibliographystyle{IEEEtran}
\bibliography{refs}

\end{document}